\documentclass[dvips,12pt]{article}
\usepackage{epsfig}
\usepackage{cite}
\def\be{\begin{equation}}
\def\ee{\end{equation}}
\def\barr{\begin{array}}
\def\earr{\end{array}}
\def\ra{\rightarrow}
\def\dis{\displaystyle}
\def\simlt{\stackrel{<}{{}_\sim}}
\def\simgt{\stackrel{>}{{}_\sim}}

\catcode`@=11 
\def \gsim{\mathrel{\mathpalette\@versim>}}
\def \lsim{\mathrel{\mathpalette\@versim<}}
\def \@versim#1#2{\lower0.4ex\vbox{\baselineskip\z@skip\lineskip\z@skip
     \lineskiplimit\z@\ialign{$\m@th#1\hfil##\hfil$%
     \crcr#2\crcr\sim\crcr}}}
\catcode`@=12 
\title{
\vspace*{-1.3cm}
\begin{flushright}
\normalsize{
ANL-HEP-PR-02-018\\
EFI-02-65\\
MRI-P-011207
  }
\end{flushright}
\vspace{1.5cm}
\Large
\textbf{ Probing Heavy Higgs Boson Models with a \\[0.3cm]
TeV Linear Collider 
} 
\vspace*{1.0cm}
\author{\large\textbf{Debajyoti Choudhury$^a$}, \textbf{T.M.P. Tait$^b$}  and 
\textbf{C.E.M.~Wagner$^{b,c}$}\\ \\
$^a$\normalsize\emph{Harish-Chandra Research Institute, Chhatnag Road,
Jhusi, Allahabad 211 019, India} \\
$^b$\normalsize\emph{HEP Division, Argonne National Laboratory,
9700 Cass Ave.,
Argonne, IL 60439, USA} \\
$^c$\normalsize\emph{Enrico Fermi Institute, Univ. of Chicago, 5640
Ellis Ave., Chicago, IL 60637, USA}}}

\begin{document}
\maketitle
\begin{abstract}
The last years have seen a great development in our understanding
of particle physics at the weak scale. Precision electroweak
observables have played a key role in this process and their values 
are consistent, within the Standard Model interpretation, with
a light Higgs boson with mass lower than about 200 GeV. If new
physics were responsible for the mechanism of electroweak
symmetry breaking, there would, quite generally, be modifications
to this prediction induced by the non-standard contributions to 
the precision electroweak observables. 
In this article, we analyze 
the experimental signatures of a heavy Higgs boson at
linear colliders. We show that a linear 
collider, with center of mass energy $\sqrt{s} \simlt 1$ TeV, would
be very useful to probe the basic ingredients of well
motivated heavy Higgs boson models: a relatively heavy 
SM-like Higgs, together with either 
extra scalar or fermionic degrees  of freedom, or with the mixing of 
the third generation quarks with non-standard heavy  quark modes.
\end{abstract}

\thispagestyle{empty}
\newpage

\section{Introduction}
\label{sec:intro}

The origin of electroweak symmetry breaking (EWSB) constitutes the
major open problem in high energy physics. It is commonly assumed
that experiments at Run II of the Tevatron collider and at the 
Large Hadron Collider (LHC) will provide the hints necessary to 
distinguish between scenarios wherein the electroweak symmetry is broken 
via strong interactions (as in Technicolor theories) as opposed to the 
case where the mechanism responsible can be described in terms of a 
perturbative theory (as in the Standard Model or its supersymmetric 
extension).  If the breakdown of the electroweak symmetry is induced 
by the formation of condensates of techniquarks via QCD-like 
interactions at energy scales of about a few TeV, one expects the 
Higgs particle to be replaced by several heavy states, including 
possibly a broad resonance, with the same quantum numbers as the 
Standard Model (SM) Higgs and a mass larger than the weak scale. 
On the contrary, supersymmetric extensions of the SM predict lighter 
Higgs boson states, with masses in the same range as those consistent 
with a SM interpretation of the precision electroweak observables.

It might be argued that the nature of the theory operating beyond the 
weak scale could be deduced (or constrained) from a study of the 
electroweak precision measurements. In supersymmetric models though,
any additional contributions to such observables progressively disappear 
as the superparticle masses increase. 
In strongly interacting models, on the other hand, the new 
contributions could very well conspire to cancel the effects due to 
the large mass of the (composite) Higgs boson in the spectrum.

{\bf A.} 
Topcolor models \cite{Hill:1991at,Bonisch:1991vd}
provide an interesting example of strongly interacting
theories which, while allowing the presence of a heavy Higgs boson state,
may still be  consistent with precision electroweak measurements. 
In the simplest top condensate models \cite{Bardeen:1990ds}, 
only the top quark is involved in the formation  of the condensate 
leading to electroweak symmetry breaking.
In this case, a top quark mass of the right order of magnitude
can be achieved only by pushing the 
cutoff scale to very large values. This naturally leads to Higgs
boson and quark masses of about 200 GeV, thereby creating a hierarchy 
problem similar to the one present in the SM. 
A supersymmetric version \cite{Clark:1990tq,Carena:1992ky} 
solves the hierarchy problem and,
for values of the compositeness scale of order of the grand unification
scale, easily accommodates the  measured top quark mass values.
In such an extension, however, a four Fermi interaction
scale of the order of the soft supersymmetry breaking masses is required,
leaving us with the problem of understanding the origin of such a scale.

Lowering the effective scale of the standard top quark four Fermi
interactions reduces the fine-tuning
problem \cite{Hill:1995hp}, 
but pushes the top quark mass to large values. 
A  solution to this problem was provided by the authors of 
Ref.~\cite{DH}. They
postulated the existence of a heavy quark, $\chi_R$ 
and its mirror partner $\chi_L$, with $\chi_R$ carrying the
same quantum numbers as the right-handed top quark. 
The new fermion $\chi$ has a large Dirac mass and the
top quark becomes light via an effective see-saw mechanism involving
the top and the $\chi$ degrees of freedom. 
For effective four-Fermi scales of about a few TeV, the model
also predicts a heavy Higgs boson state. This does not lead
to a conflict with electroweak precision observables 
since the  mixing of the top quark with the heavy state induces modifications
to the $\rho$ parameter, which counteract the effects of the heavy
Higgs boson present in the spectrum \cite{Chivukula:1999wd}.

Since the Dirac mass of the heavy state tends to be quite large
(about of 3 to 10 TeV), and the effective theory at energies
below this mass is a SM one with a heavy Higgs, it
is obvious that only a very large hadron collider would
be able to directly probe the fundamental degrees of
freedom \cite{Barklow:2002su}. In this work we show, that by
detecting a heavy Higgs and measuring the
precise value of the top quark coupling to the $Z$
gauge boson, a TeV Linear Collider can provide the means
to analyze these models and even predict the presence of a heavy 
fermion in the spectrum by establishing its mixing with the top.

{\bf B.} Two Higgs doublet models provide the simplest extension of the SM, 
and arise naturally in extensions of the 
top-seesaw scenario described above \cite{Collins:2000rz,He:2001fz}. 
As in the SM, there is generically no prediction for the values
of the charged, CP-even and CP-odd Higgs bosons (for simplicity we
shall assume that there is no violation of CP in the Higgs effective
potential\footnote{For the case of CP-violation see
Ref.~\cite{Carena:2000yi} 
and references therein.}). 
Constraints on the Higgs boson masses can still be
achieved from the comparison of the two Higgs doublet predictions
with the measured values of the precision electroweak observables.
However, while in the SM the Higgs boson needs to be
light, no such constraint exists in the two Higgs doublet models.
The reason is that the extra degrees of freedom can provide
the contributions necessary to compensate for the heavy Higgs boson.
This is achieved by splitting the components of the effective non-standard
Higgs doublet. Therefore, one can easily imagine a scenario in which
all the Higgs bosons are heavy without implying any conflict with
the precision electroweak observables~\cite{Chankowski:2000an}. 
However, the heavier the SM-like 
Higgs boson, the larger is the effective quartic coupling
leading to its mass. Moreover, for a fixed given value of 
the non-standard contribution to the precision electroweak observables,
the heavier the non-standard
Higgs boson, the larger the necessary splitting of masses, induced by 
additional quartic couplings, needs to be. Therefore, a heavy Higgs spectrum,
with only a heavy SM-like Higgs at the reach of TeV
Linear Collider would imply the presence of very large quartic couplings. 
In this article we show that requiring perturbative consistency of
the theory up to a scale of order of a few TeV demands the presence of
at least one SM-like Higgs boson with mass lower than 600 GeV. Moreover, 
for a SM-like Higgs boson mass larger than 450 GeV, extra non-standard
scalar degrees of freedom should appear at a linear collier.

{\bf C.} The mixing of vector-like fermions with the SM quarks 
can produce important modifications to the
precision electroweak parameters if they mix strongly with the third
generation quarks. Due to the strong mixing, this  
could happen even if these extra fermions are relatively light. An
example of this case are the Beautiful Mirrors, recently 
proposed~\cite{Choudhury:2001hs}
by the authors to resolve the discrepancy between the measured
forward-backward asymmetry of bottom quarks and its SM prediction.
In one of the two cases studied in Ref.~\cite{Choudhury:2001hs}, 
the fit to the precision electroweak data is significantly
improved when there is a heavy Higgs boson in the spectrum. 
This further implies that the new quarks must be lighter than
roughly 300 GeV, rendering them easily observable at a Linear Collider,
and allowing a measurement of the mixing angle between standard and mirror
bottom quarks through the flavor-violating process in which one
light and one heavy state are produced.

This article is organized as follows. In section~\ref{sec:hhiggs}, 
we review the experimental signatures of a heavy Higgs boson with SM-like
properties. This implies, for example, that it decays predominantly
into pairs of $W$, $Z$, and top.  A heavy SM-like Higgs need not be
a feature of {\em any} theory of physics beyond the Standard Model,
but is approximately realized in the three models we proceed to 
analyze in detail.  For definiteness, in the numerical simulations,
we treat the Higgs as having precisely the properties of a heavy Higgs
in the SM.  We show that, for the mass range under consideration,
the Higgs is a broad resonance and an accurate description of its
signatures forbids the decoupling of the production and decay processes
(the narrow width approximation). An estimate of the reach
of a $\sqrt{s} = 800$ GeV collider is given. In section~\ref{sec:topcolor}, we
analyze the means to probe Top Seesaw Models. We show how the
interplay of a Higgs search with an accurate measurement of the 
top quark coupling can shed light on the properties of this model
and even predict the mass of a heavy quark with the same quantum numbers
as the right-handed top quark. In section~\ref{sec:2hdm}, we discuss 
the bounds on the two Higgs doublet model arising from the requirement 
of perturbative consistency of the theory up to a scale of the order of at
least a few TeV combined with consistency with electroweak precision
data. In section~\ref{sec:bmirrors}, we discuss the Beautiful Mirrors
at a linear collider, and demonstrate that one can measure the
mixing between the heavy vector-like quarks and the bottom quark. 
We reserve section~\ref{sec:conclusions} for our conclusions.

\section{Heavy Higgs Boson}
\label{sec:hhiggs}

Since our concern is with theories containing heavy Higgs bosons,
we begin by considering how well a heavy SM-like Higgs can
be studied at a linear collider.  This question is of interest
in itself because, in the SM, a Higgs with mass greater than
about 350 GeV has an appreciable width: for $m_H = 350$ GeV,
the width  $\Gamma_H$ is roughly = 10 GeV, and by 
$m_H = 600$ GeV, the width is more than 100 GeV.  
The reason for this large enhancement of the width 
lies in the origin of the longitudinal modes of the $W$ and $Z$ bosons.
For the partial decay width of the a scalar Higgs boson 
into a pair of on-shell weak gauge bosons, we have
\be
\Gamma(H \rightarrow VV) \simeq  \frac{G_F (|Q_V| + 1)} 
{\sqrt{2} \, 16 \pi}  m_H^3 \ 
\left( 1 - \frac{4 M_V^2}{m_H^2} + 3 \frac{4 M_V^4}{m_H^4} \right)  \ 
\left( 1 - \frac{4 M_V^2}{m_H^2} \right)^{1/2}
\ee
where $Q_V$ is the charge of the $W$ ($Z$) while $M_V$ is its mass. 
Therefore, for large Higgs boson masses,  the width grows as $m_H^3$. 
This behavior can be easily understood from the equivalence theorem
\cite{Cornwall:1974km,Lee:1977eg,Chanowitz:1985hj}, which dictates that,
in high energy processes, weak bosons may be replaced by the corresponding
pseudo-Goldstone bosons (pGB's).  The Higgs coupling to two pGB's
is given by the Higgs quartic interaction
times the vacuum expectation value (VEV), $\lambda v$.  Thus, the
partial decay width into weak bosons is proportional (for large
Higgs masses) to $(\lambda v)^2 / m_H$, where $m_H$ accounts for the 
phase space of the decay.  Using the tree-level relation
between the Higgs mass and $\lambda$, namely $m_H^2 = 2 \lambda v^2$,
one then finds the partial width into weak bosons to be proportional to
$m_H^3 / v^2$.

This should be compared to the partial decay widths into fermions, 
which are only linear in the Higgs mass.
\begin{equation}
\Gamma(H \to f \bar{f}) \simeq 
\frac{N_c G_F m_f^2}{\sqrt{2} \, 4 \pi}  m_H
\left( 1  - \frac{4 m_f^2}{m_H^2} \right)^{3/2}
\end{equation} 
At these large
masses, $m_H \simgt 400$ GeV, 
the decay is about $60\%$ into $W$ pairs, $20\%$ into $Z$
pairs, and the remainder into $t \bar t$ pairs\footnote{These 
results conform with those obtained from HDECAY~\protect\cite{Spira:1997if}.}. 

The simple picture of the Higgs as a particle that is produced
with some cross section (for example, as $Z H$), and then decays 
into something or other breaks down at large Higgs masses. 
First, the large width means that, in order to study the production 
of a Higgs and its subsequent decay products, one should take into
account interference effects between contributions involving a Higgs
boson and  the continuum production of
the same decay products through diagrams that do not directly involve 
a Higgs. Indeed,  there is no meaningful way to separate the Higgs 
``signal'' from the
continuum ``background'' as is commonly done for narrow resonances
and these interference effects may be quite significant. Moreover,
approximating the Higgs signatures by a convolution of production and
decay processes may lead to quantitatively bad estimates,
particularly when the center of mass energy is close to the Higgs
production threshold.
The situation is further complicated in that the Higgs plays a key role
in unitarizing the triple- and quadruple- gauge boson interactions at
high energies, and thus the background ``without the Higgs'' is 
inconsistent.  However, we are not sensitive to this problem, because
our initial state consists of massless leptons which force some of the
internal vector bosons to be transversely polarized, for which
unitarity violations do not occur \cite{Lee:1977eg}.

Our interest is in the case in which the heavy Higgs is SM-like, and
we would like to see how heavy the Higgs might be and yet still be
observed and identified as such.
Hence we define the background to Higgs production 
to be that emanating from 
the full (interfering) set of resonant and non-resonant production
graphs in the limit of large Higgs mass, $m_H \ra 1$ TeV.
We have verified that our results for the background do not differ
significantly if we choose heavier Higgs masses.
This is justified given the fact that our specific observables do not 
suffer from unitarity violations.
In fact, because the unitarization of longitudinal gauge boson scattering
occurs order-by-order in perturbation theory, one encounters problems with 
the Breit-Wigner prescription for the form of the propagator, 
which sums Higgs self-energy contributions at all orders.
We have used the modified Higgs propagator proposed in
\cite{Seymour:1995qg} (see also \cite{Basdevant:1992nb}), 
which circumvents this problem, and observe
that for our results, there is essentially no difference between the
modified propagator and the naive Breit-Wigner.

The large branching ratio of $H \ra W^+ W^-$ implies that polarization
of the initial electron and/or positron beams can be very effective
at reducing backgrounds to heavy Higgs searches, since it suppresses
radiation of the $W$ bosons from the $e^-$ or $e^+$ lines, and can
reduce these backgrounds by roughly an order of magnitude.  Some of the
Higgs production modes, such as $W$-fusion modes, will also be significantly
affected by polarization, so the question as to whether or not polarization
is beneficial depends on the production and decay mode of the Higgs
under study.  

\begin{figure}[htb]
\centerline{
\epsfxsize=18cm \epsfysize=10.0cm
                     \epsfbox{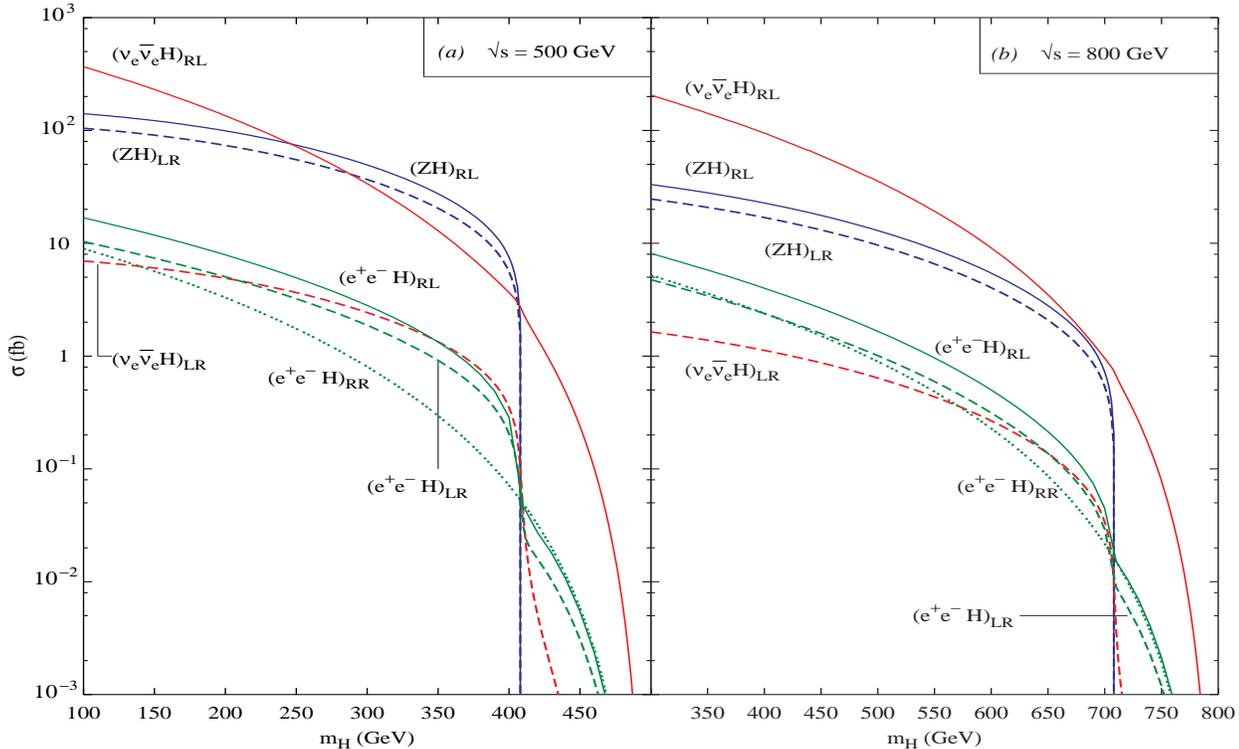}
}
\caption{\em Cross sections for Higgs production as a function of Higgs
mass through three different processes at an $e^+ e^-$ collider with 
$\sqrt{S}=500 (800)$ GeV.  The ordered subscripts ($LR$, $RL$ and $RR$)
refer to the (100\%) polarizations of the $e^+$ and $e^-$ respectively.}
\label{fig:production}
\end{figure}

In order to properly analyze the Higgs signatures, it
is useful to analyze the naive  Higgs production
cross sections which, as explained above, can only serve
as an indicator of the real production cross section to be
obtained when full width effects are taken in to account.
In Fig.~\ref{fig:production}
we show the Higgs production cross sections in the modes:
\begin{itemize}
\item $e^+ e^- \ra Z H$ ($Z H$ production)
\item $e^+ e^- \ra \nu_e \bar{\nu}_e H$ ($W$-fusion)
\item $e^+ e^- \ra e^+ e^- H$ ($Z$-fusion)
\end{itemize}
for a linear collider of $\sqrt{S}= 500$ GeV and 800 GeV, respectively.
To demonstrate the effect of beam polarization, we have chosen the 
ideal case of 100\% polarization (right- or left-handed) for both 
the positron and the electron. The tiny electron mass
ensures that only two initial state polarization combinations 
may contribute in each of the first two cases above. 
Results for partially polarized beams may easily 
be obtained by forming an appropriate
weighted sum of the two polarization combinations shown.

Note that the $W$-fusion process tends to be the dominant process with
 $Z$-fusion being relatively unimportant. While beam polarization is 
not very important in $Z H$ production, it certainly plays a crucial 
role in the other two, particularly for $W$-fusion. This is only to 
be expected as the $e_L^+ e_R^-$ initial state cannot support $W$-fusion. 
Hence, lack of beam polarization would only serve to 
reduce the total $W$-fusion 
cross section by an approximate factor of 4.

\begin{figure}[ht!]
\vspace*{-6ex}
\centerline{
\epsfxsize=9.4cm \epsfysize=9.5cm 
                     \epsfbox{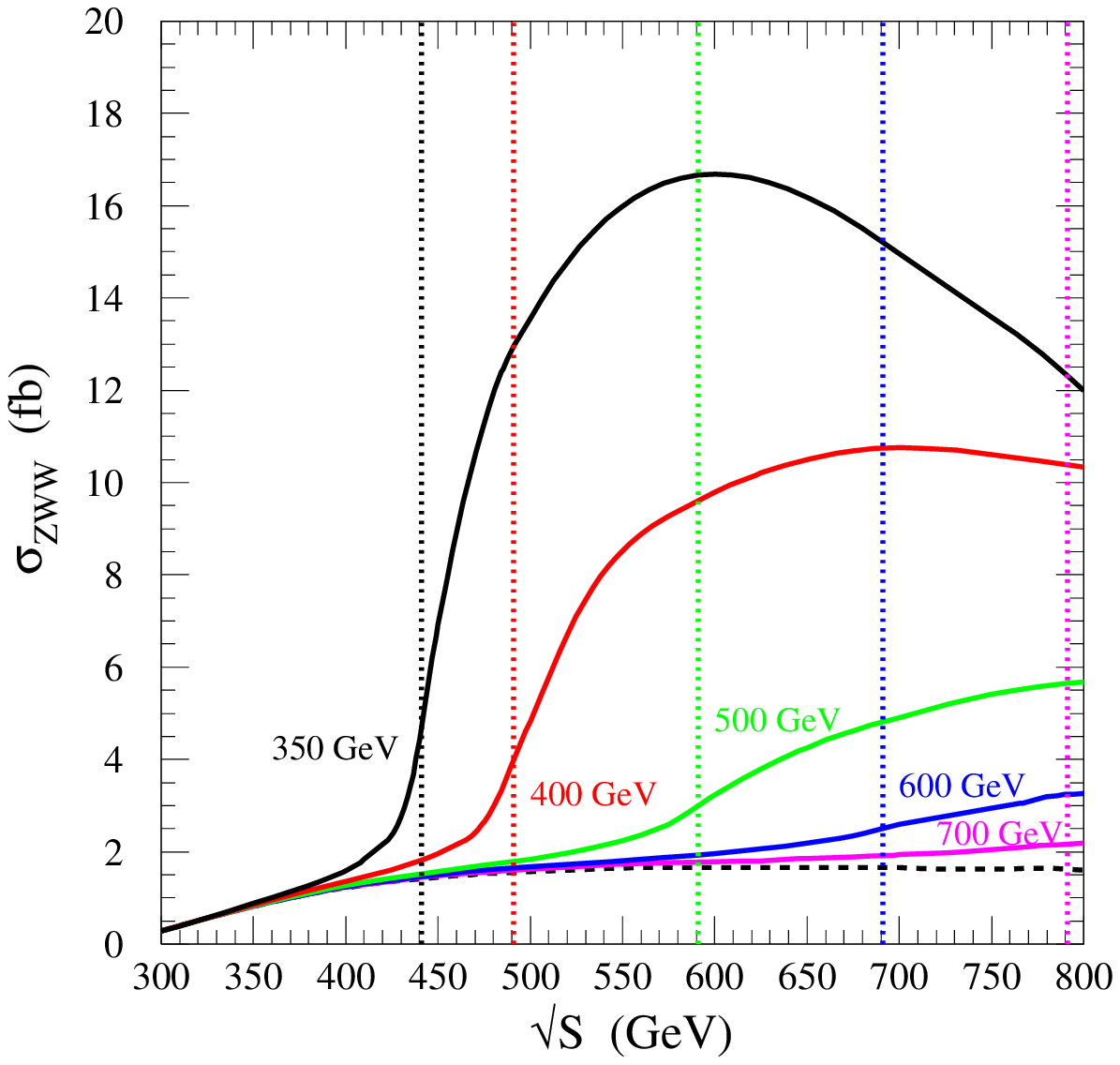}
        \hspace*{-8ex}
\epsfxsize = 9.4cm \epsfysize=9.5cm 
        \epsfbox{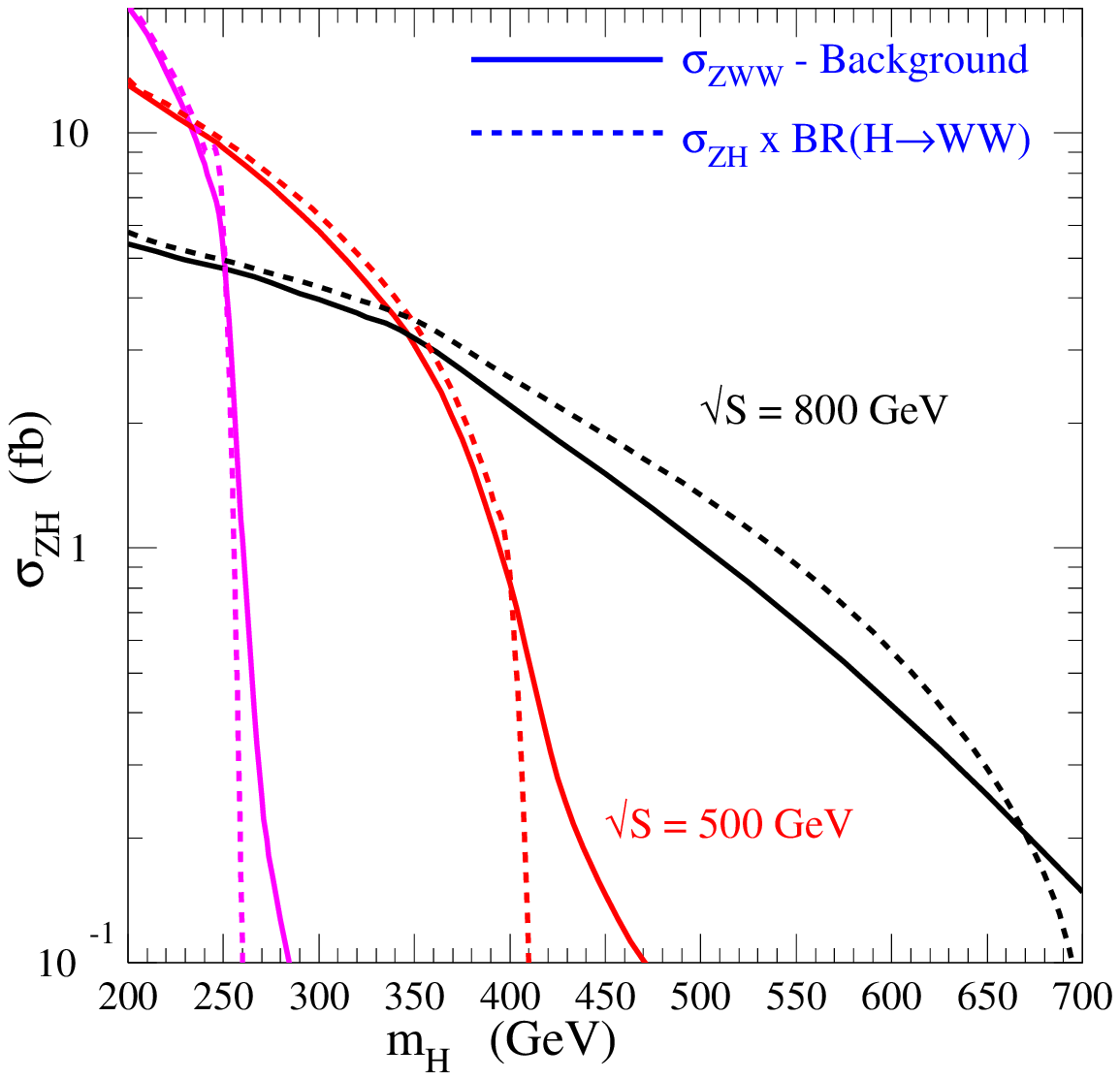}
}
\vspace*{-3ex}
\caption{\em Cross sections for $Z W^+ W^-$ production as  a function of
{\em (a)} $\sqrt{S}$ for a variety of Higgs masses at a linear collider,
assuming $RL$ polarization of the $e^-$ ($e^+$). The dotted lines
corresponding to each curve show the kinematic limit for on-shell
Higgs production, and the dashed line indicates the cross section
when the Higgs mass is very large, and is thus
indicative of the background; {\em (b)} as a function of the Higgs mass for
different $\sqrt{S}$. The solid and dashed line denote the full 
three body process simulation
and the two body production plus Higgs decay approximation, respectively.}
\label{fig:zww}
\end{figure}

We proceed now to examine
the influence of the large Higgs width on these naive
``on-shell'' production curves in the context of specific
decay modes of the Higgs.
This forces us to include the non-resonant graphs that would be
associated with the background in a narrow resonance search.
We begin with $e^+ e^- \ra Z^* \ra Z H \ra Z W^+ W^-$ which we
expect from Fig.~\ref{fig:production}
to be an important search mode up to the kinematic limit of
$\sqrt{s} - M_Z \geq M_H$.  Clearly, this mode benefits from
$RL$ polarization of the $e^-$ ($e^+$) which reduces the cross section
only modestly, while significantly removing a large portion of the
$Z W^+ W^-$ non-resonant production.  In Fig.~\ref{fig:zww}, we
show the production cross section for $RL$ polarization as a
function of the collider energy, and for a variety of Higgs masses.
For reference, the kinematic limits on production are also shown.
The effect of the Higgs width, especially for masses above 500
GeV, is striking.  The Higgs contributions begin much earlier
than the kinematic limit imposed by the on-shell Higgs approximation.
In Fig.~\ref{fig:zww}, we also show the $ZH$ production cross section
as a function of the Higgs mass for different center of mass energy. 
As is 
evident from the figure, 
above the kinematic limit, the on-shell approximation over-estimates
the cross section. Higgs production, however, extends well beyond
what is expected from the naive on-shell approximation, and
therefore, for sufficient luminosity, the Higgs reach may be
actually underestimated compared to the real reach once width effects
are taken into account.

\begin{figure}[ht!]
\vspace*{-3.7cm}
\centerline{
\epsfxsize=10.0cm \epsfysize=11cm 
                     \epsfbox{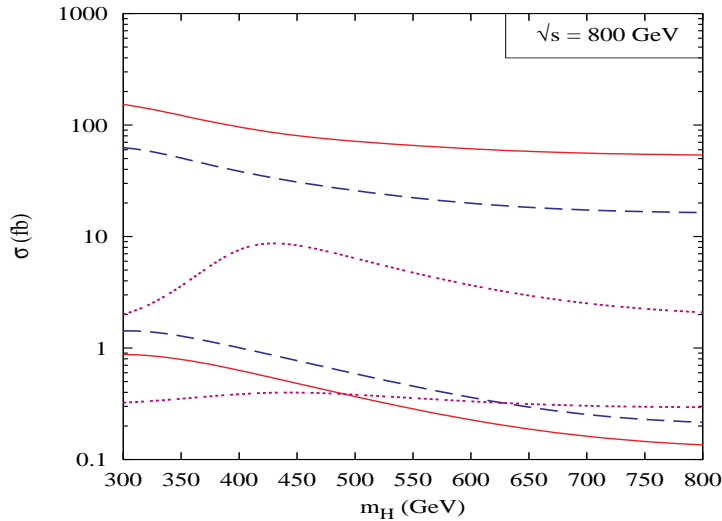}
}
\caption{\em Cross sections for $e^+ e^- \ra \nu_e \bar \nu_e W^+ W^-$ 
	(solid), $e^+ e^- \ra \nu_e \bar \nu_e Z Z$ (dashed) and
	 $e^+ e^- \ra \nu_e \bar \nu_e t \bar t$ (dotted) as a function
	of the Higgs mass. The upper and lower sets correspond to 
	($e^+_R, e^-_L$) and ($e^+_L, e^-_R$) initial states 
	respectively.}
\label{fig:wfusion}
\end{figure}

A very similar story holds for the other dominant Higgs production process,
{\em viz.} $WW$ fusion. In Fig.~\ref{fig:wfusion} we present the cross 
sections for the three major final states emanating from a heavy Higgs 
thus produced. Although the cross sections may seem to be significant
vis a vis the projected luminosities at a linear collider, the 
continuum background is very large too. So while discovering a relatively 
light Higgs may would be easy, the same is certainly not true for 
$m_H \gsim 550$~GeV. It is instructive to look at the relative deviations
in the cross sections caused by the presence of the scalar. For the 
$\nu_e \bar \nu_e W^+ W^-$ and the $\nu_e \bar \nu_e Z Z$ final states,
a very large fraction of the cross section arises from non-resonant 
diagrams. Consequently, this deviation is more pronounced for an initial 
state starting with the `wrong' polarization ($e^+_L, e^-_R$). 
Whether  this advantage overcomes the drawback of much smaller statistics 
is of course a matter of a quantitative analysis.

As the preceding discussion demonstrates, 
the question as to how heavy a Higgs is still amenable
to be discovered at a high energy linear collider 
cannot be answered trivially. 
Even a semi-realistic assessment of the experimental efficiency 
needs detailed simulations of the events. 
While such a study is beyond the scope of this work,
it is interesting nevertheless to examine the event kinematics.
We focus on the kinematics of the $W$-fusion process (with $RL$ polarization),
as it represents, for fixed collider energy, the largest production
mode for the highest Higgs masses, and is thus expected to provide
the best hope of discovery for a very heavy Higgs.
Apart from indicating possible ways to enhance the signal to 
background ratio, this will also allow us to gain some insight into 
the importance of the large width and its impact on the observability 
of a Higgs with a large mass and (approximately) SM-like interactions.  

\begin{figure}[t]
\vspace*{-3cm}
\centerline{
\epsfxsize=8.0cm \epsfysize=10.0cm 
                     \epsfbox{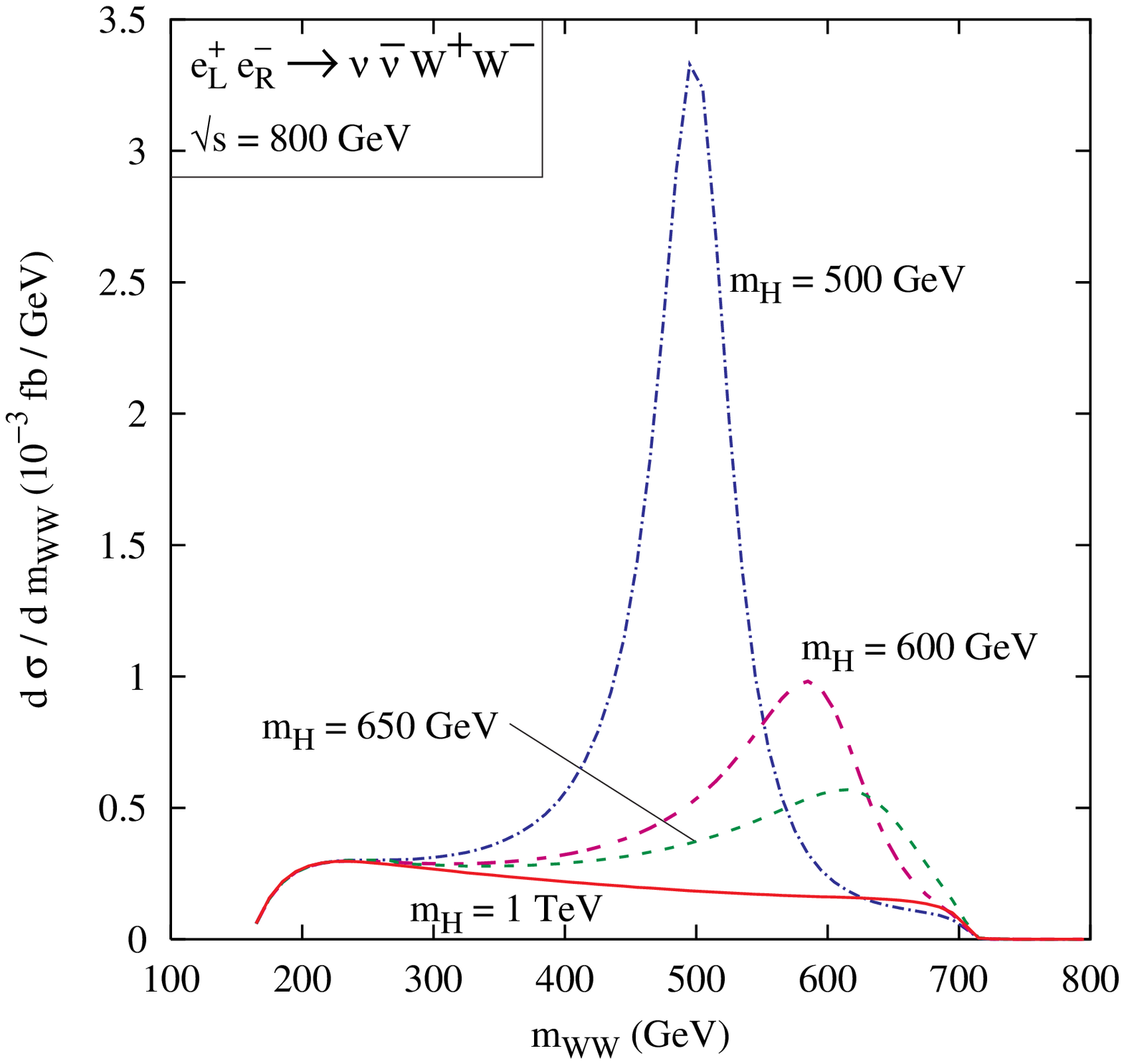}
        \hspace*{-2ex}
\epsfxsize = 8.0cm \epsfysize=10.0cm 
        \epsfbox{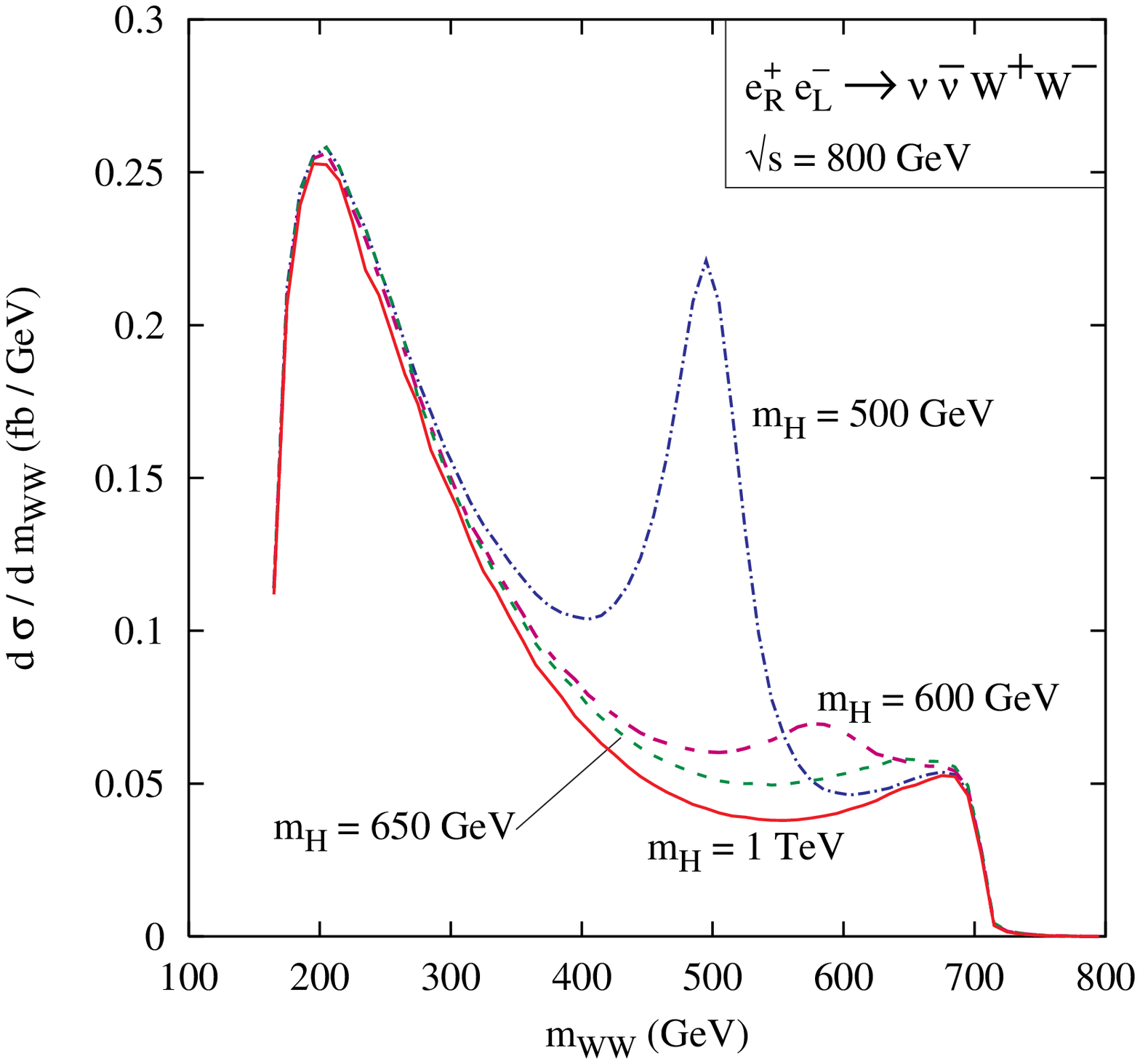}
}
\vspace*{-18ex}
\centerline{
\epsfxsize=8.0cm \epsfysize=10.0cm 
                     \epsfbox{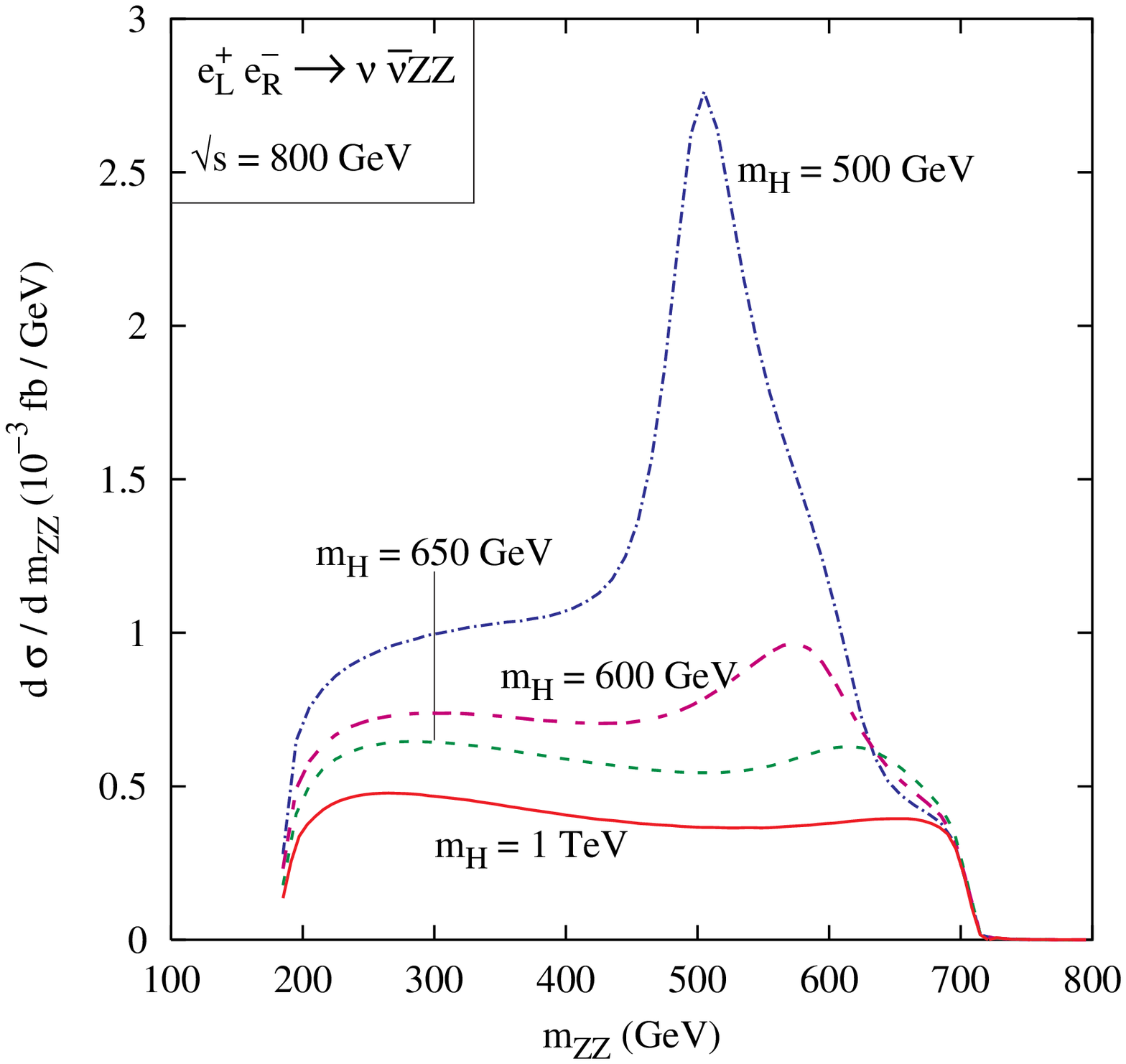}
        \hspace*{-2ex}
\epsfxsize = 8.0cm \epsfysize=10.0cm 
        \epsfbox{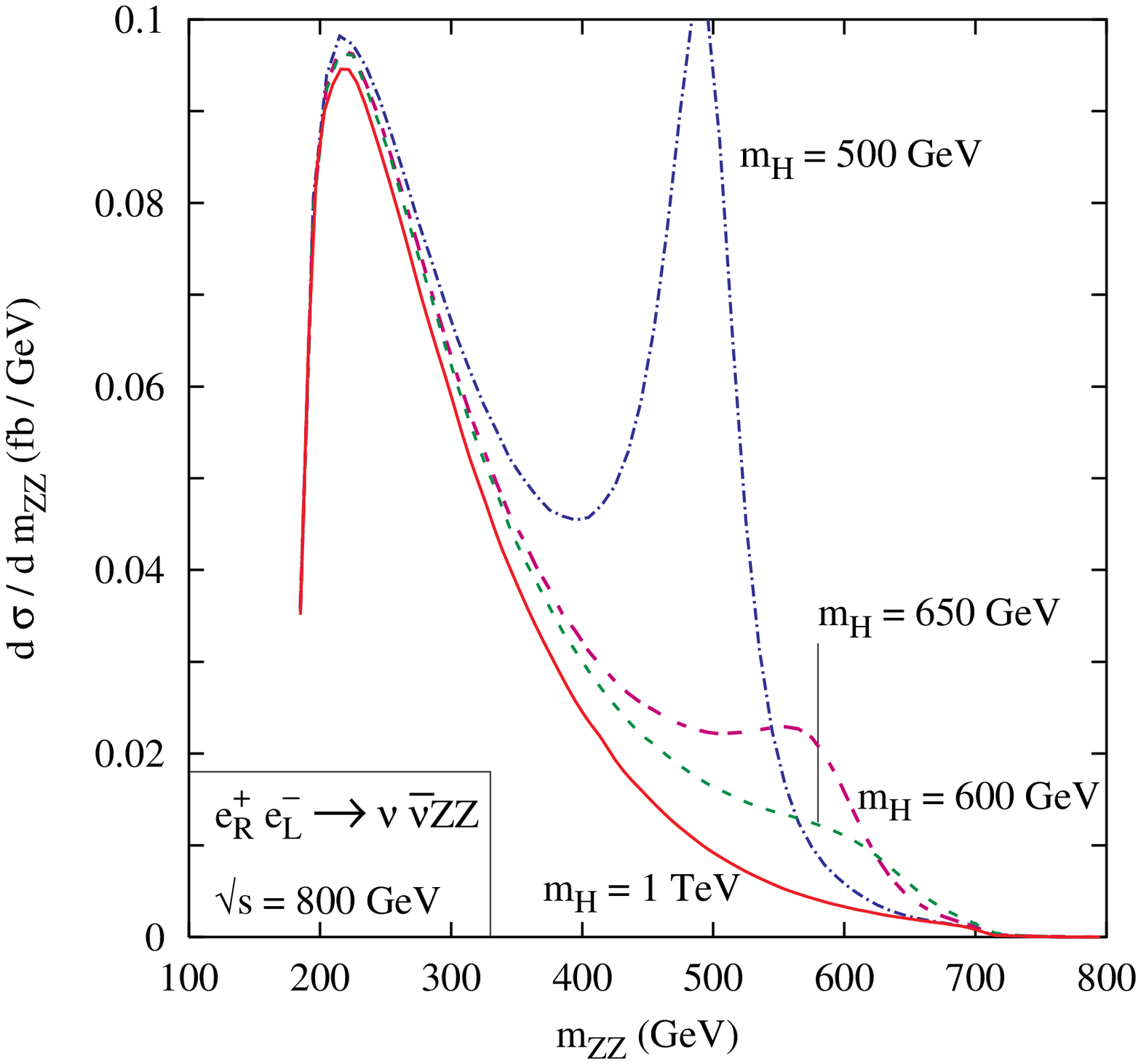}
}
\vspace*{-18ex}
\centerline{
\epsfxsize=8.0cm \epsfysize=10.0cm 
                     \epsfbox{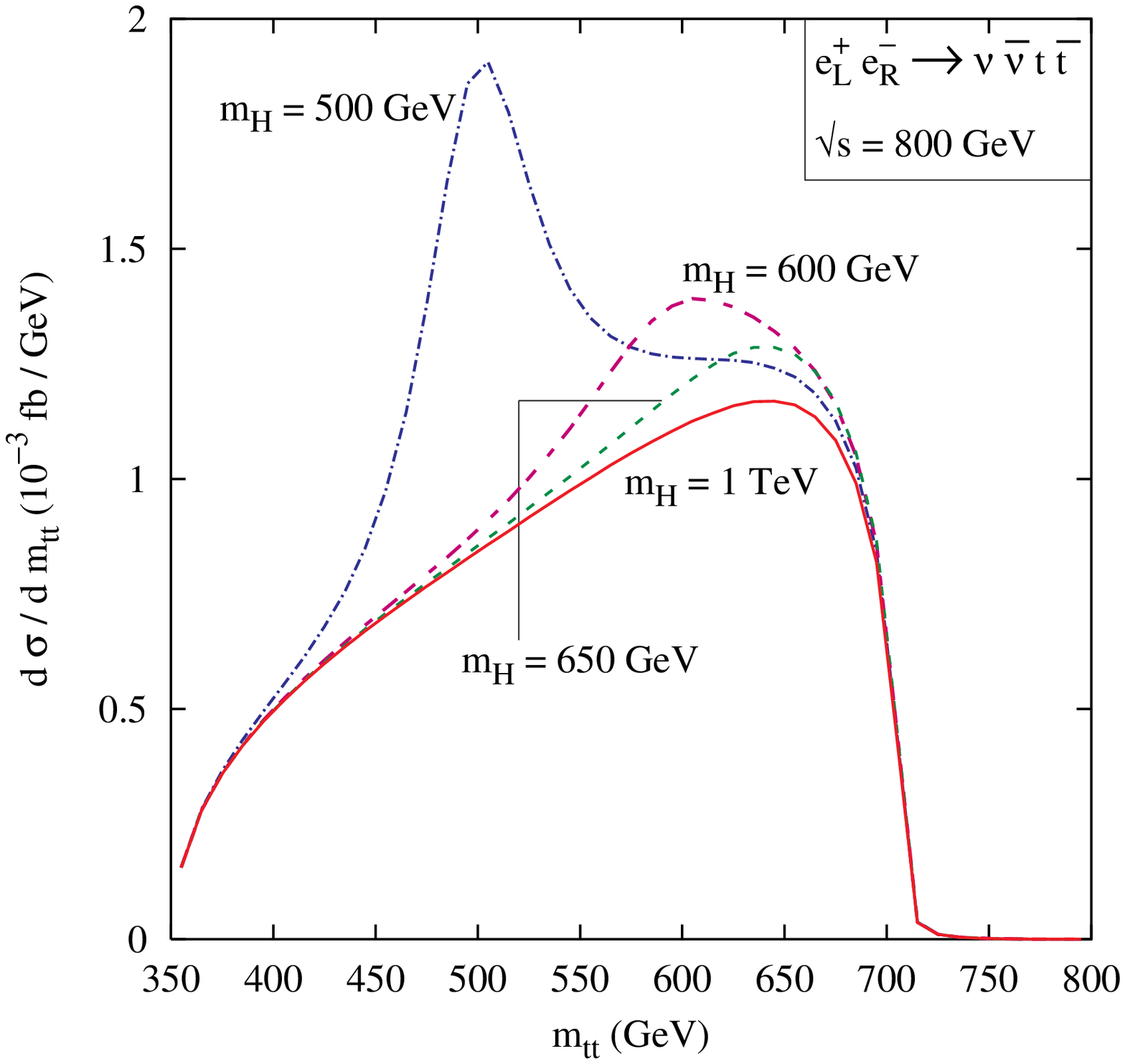}
        \hspace*{-2ex}
\epsfxsize = 8.0cm \epsfysize=10.0cm 
        \epsfbox{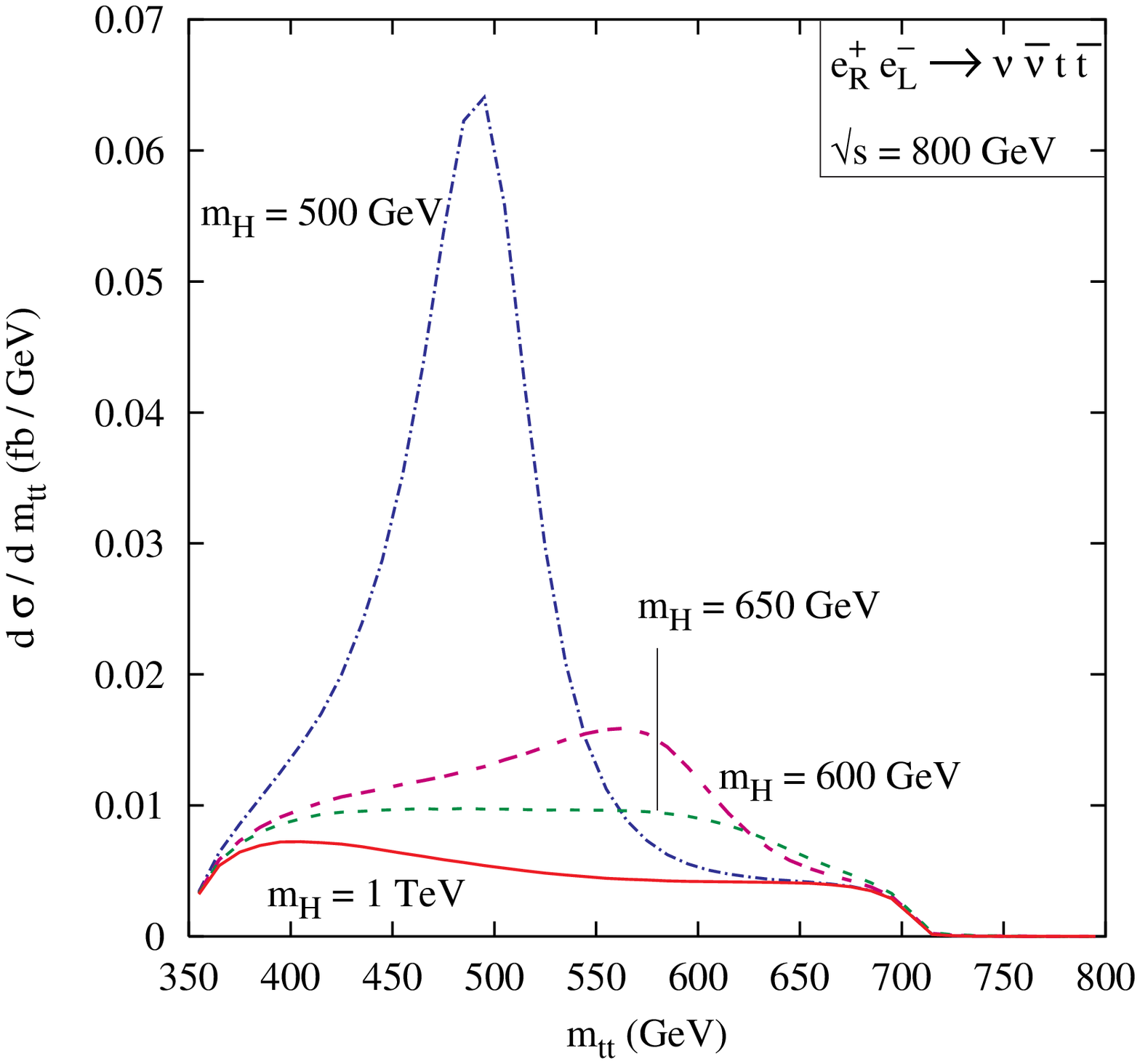}
}
\caption{\em Invariant mass distribution of the `visible' part
in the processes 
$e^+ e^- \ra \nu \overline{\nu} W^+ W^-$,  $\nu \overline{\nu} Z Z$
and $\nu \overline{\nu} t \overline{t}$
in the presence of a Higgs. Three representative masses are chosen.
Also shown are the predictions in the ``no Higgs'' limit ($m_H = 1$ TeV).
For each case, two 
different choices of positron and electron polarizations 
have been used. 
}
\label{fig:mwwRL}
\end{figure}

While angular distributions of the final state particles 
do bear the mark of a scalar, we find that 
the invariant mass distribution
of the Higgs `decay products' is perhaps the best discriminant.
This is not unexpected. Rather, this 
is the very variable that one would have concentrated on for a 
resonant production process.
In Figs.~\ref{fig:mwwRL}, we present the corresponding distributions
for each of the processes considered in Fig.~\ref{fig:wfusion}.
We note that polarization
can widely influence the signal to background ratio, and thus can play
an essential role in heavy Higgs studies at the LC.  We further see a general
trend that while a clear resonant peak is obtained for a Higgs mass of 500
GeV, for 600 GeV it is much smaller and wider. And for 650 GeV, a
peak is almost non-existent with the excess spread out almost evenly. 

In Table~\ref{table} we show the signal and background cross sections
associated with different final states consistent with the 
weak-fusion production of a heavy Higgs at an 800 GeV linear collider.
The optimal mass window for the discovery of such a heavy Higgs is
displayed and the luminosity required for $3 \sigma$ evidence or a
$5 \sigma$ discovery of the heavy Higgs boson is displayed for 
different choices of the electron and positron polarization. 
We stress again that careful study,
including efficiencies and modeling of detector effects, is
needed to draw firm conclusions, but it seems reasonable to expect that
a SM-like Higgs with mass below 650 GeV could be
identified in its principle decay modes from $100~{\rm fb}^{-1}$
of data at a polarized 800 GeV linear collider.
This is to be compared with the expected discovery reach of up to 
1 TeV by the LHC \cite{Mitsou:2000nv}.  Of course, once
the Higgs has been discovered at the LHC, it will be up to the linear 
collider to carefully study its properties, and thus the question of 
its observability there remains of the utmost importance.

\clearpage
\begin{table}[htb]
\label{table}
\begin{tabular}{| l | c | c | r | r | r | r | r | r |}
\hline
& \multicolumn{1}{c|}{Final state} 
&  \multicolumn{1}{c|}{$(P_{e^+}, P_{e^-})$}
& \multicolumn{1}{c|}{$m_H$}
& \multicolumn{1}{c|}{Mass window}
& \multicolumn{1}{c|}{Signal}  
& \multicolumn{1}{c|}{Bkgd.}
& \multicolumn{2}{c|}{Reqd. lumin.}\\
& & & (GeV) & \multicolumn{1}{c|}{(GeV)}
                    & \multicolumn{1}{c|}{(fb)} 
                    & \multicolumn{1}{c|}{(fb)} 
	            & \multicolumn{1}{p{1cm}|}{${\cal L}_{3 \sigma}$ (fb$^{-1}$)}
                    & \multicolumn{1}{p{1cm}|}{${\cal L}_{5 \sigma}$ (fb$^{-1}$)}\\
\hline
1. & $\nu \bar \nu W^- W^+$ & (R, L) 
		& 500 & (450, 530) & 11.3 & 3.8 &  0.34 & 1.2 \\
   & & 
		& 600 & (440, 620) & 4.2  &  8.0 & 4 & 11 \\
   & & 
		& 650 & (390, 660) & 2.9  & 13.1 & 14 & 39 \\
\cline{3-9}
   &  & (L, R) 
		& 500 & (450, 530) & 0.195 & 0.017 &  10 & 33 \\
   & & 
		& 600 & (490, 630) & 0.087  & 0.025 & 37 & 144 \\
   & & 
		& 650 & (490, 670) & 0.056  & 0.031 & 86 & 240 \\
\hline
2. & $\nu \bar \nu Z Z$ & (R, L) 
		& 500 & (450, 540) & 5.9 & 0.96 &  0.44 & 1.4 \\
   & & 
		& 600 & (500, 650) & 1.8  &  0.72 & 2.4 & 8.4 \\
   & & 
		& 650 & (500, 740) & 1.0  & 0.78 & 6.5 & 22.3 \\
\cline{3-9}
   &  & (L, R) 
		& 500 & (440, 590) & 0.243 & 0.059 &  14 & 43 \\
   & & 
		& 600 & (180, 640) & 0.143  & 0.188 & 82 & 228 \\
   & & 
		& 650 & (180, 680) & 0.075  & 0.204 & 325 & 902 \\
\hline
3. & $\nu \bar \nu t \bar t$ & (R, L) 
		& 500 & (450, 530) & 3.5 & 0.49 &  0.51 & 2.26 \\
   & & 
		& 600 & (460, 620) & 1.3  &  0.81 & 4.8 & 15.6 \\
   & & 
		& 650 & (440, 660) & 0.82  & 1.1 & 14.9 & 41.4 \\
\cline{3-9}
   &  & (L, R) 
		& 500 & (460, 540) & 0.065 & 0.077 &  165 & 458 \\
   & & 
		& 600 & (530, 670) & 0.024  & 0.163 & 2420 & 6710 \\
   & & 
		& 650 & (570, 720) & 0.009  & 0.147 & 16200 & 45100 \\
\hline
\end{tabular}
	\caption{\em The optimal mass windows that maximize $S / \sqrt{B}$ 
		for particular final states emanating from a given 
		polarization combination for the $e^+ e^-$ pair. Also 
		shown are the luminosities required for obtaining a 
		``$3 \sigma$''
		and a ``$5 \sigma$'' excess within the mass window. 
		Within the given window, a 100\% efficiency is assumed. }
\end{table}

\section{Top Seesaw Model}
\label{sec:topcolor}

We now study a promising model of dynamical EWSB with a heavy
Higgs that can be consistent with precision electroweak data:
the Top Seesaw.  Its simplest implementation starts with the
gauge group $SU(3)_1 \times SU(3)_2$, which is broken to 
$SU(3)_c$ (QCD) by the introduction of a bi-fundamental field $\Phi$, which 
parameterizes the unspecified Topcolor-breaking dynamics. The left-handed
component of the top quark transforms in the fundamental representation
of $SU(3)_1$, while the right-handed component transforms in the
fundamental representation of $SU(3)_2$. 
An additional singlet fermion $\chi$ has left- and right-handed
components which transform in the fundamental
representation of $SU(3)_2$ and $SU(3)_1$, respectively. After the breakdown
of the strong gauge groups to $SU(3)_c$, mass terms involving
the left-handed component of $\chi$ with the right handed components
of $\chi$ and $t$ are allowed. 
We shall called these mass terms
$m_{\chi\chi}$ and $m_{\chi t}$ respectively, and they are naturally
of order of the scale of breaking of the strong gauge groups. 
Only one Dirac fermion acquires mass, with left handed component
given by $\chi_L$ and right handed component given by a combination
of $\chi_R$ and $t_R$. 

The massive strong vector bosons (colorons) induce four-Fermi interactions
involving the $t_L$ and $\chi_R$ currents. 
There are expected to be other interactions, but for natural values
of $m_{\chi t}$, they do not lead to any relevant phenomena in the
process of electroweak symmetry breaking. After a Fierz transformation,
the relevant interaction reads
\begin{equation}
{\cal{L}}_{int} \simeq \frac{g^2}{M^2} \left(\bar{\psi}_L \chi_R \right)
\left( \bar{\chi}_R \psi_L \right)
\end{equation}
where $g$ is the strong Topcolor interaction strength, 
$M$ is the mass of the heavy coloron gauge boson, and $\psi_L = (t,b)_L$.
For appropriate values of $\kappa = g^2/ 4 \pi$, 
the model undergoes a spontaneous breakdown of the electroweak symmetry.
In the massless limit, $m_{\chi \chi}, m_{\chi t} \ra 0$, 
the  Nambu-Jona-Lasinio (NJL) critical coupling for spontaneous 
symmetry breaking \cite{NJL} is given by $\kappa_c = 2 \pi / N_c$.  
The effect of nonzero $m_{\chi \chi}$ and $m_{\chi t}$ is to require
a somewhat larger coupling for EWSB \cite{He:2001fz}.
However, taking the Topcolor coupling arbitrarily strong represents
fine-tuning, in the sense that one is forcing two a priori unrelated
quantities (the Topcolor breaking scale $M$ and the scale at which
Topcolor would have confined had it remained unbroken) to
be very close to one another.  Thus it is more natural to assume
that the Topcolor interaction is only slightly super-critical.
For definiteness, we will confine our discussion to 
$\kappa / \kappa_c < 1.2$

The immediate effect of the dynamical symmetry breaking is to
induce a mass $\mu_{t \chi}$ mixing the
left-handed component of $t$ with the right handed component of
$\chi$. The diagonalization of the fermion mass matrix leads
to two mass eigenstates; a heavy one of mass of order $m_{\chi\chi}$
and a light one, which should be identified with the top quark.
In order to distinguish the interaction eigenstates from the mass
eigenstates, we henceforth denote the interaction eigenstates
as primed fields, i.e. $t^\prime$, and the physical (mass) eigenstate
fields as unprimed.

The mass parameter $\mu_{t \chi}$ required to generate appropriate
$W$ and $Z$ masses is approximately given by the
value that the top quark mass would acquire in the absence of 
the fermion $\chi$, starting with a large Yukawa coupling value
at the scale $\Lambda \simeq {\cal{O}}(M)$. 
This value is of order 700 GeV for 
$\Lambda$ of order of a few TeV. For $m_{\chi\chi} \gg m_{\chi t}
\gg \mu_{t \chi}$, the top quark mass is given by 
\begin{equation}
m_t \simeq \frac{\mu_{t \chi} m_{\chi t}}{m_{\chi \chi}} .
\end{equation}
It is clear that the value of the top quark mass
defines the ratio of the parameters $\mu_{t \chi}$ and $m_{\chi \chi}$.
A detailed analysis of the parameter space of 
$\Lambda$, $m_{\chi \chi}$, and $m_{\chi t}$ which provide acceptable
$W$, $Z$, and $t$ masses may be found in \cite{He:2001fz}.
These parameters also determine 
the mixing of the right- and left-handed components of the top
and the heavy quark.  Because $t^\prime_L$ 
is a component of a weak doublet while $\chi^\prime_L$ is a weak singlet, 
the left-mixing has a direct influence on the interaction of the physical 
top with the weak bosons.  On the other hand, the right-handed mixing
is between two states with the same quantum numbers, and thus has
no impact on the coupling to the gauge fields\footnote{The right-handed
mixing does have a slight impact on the coupling with the composite
Higgs boson(s), though generally the deviation is at the level of per cent,
too small to measure at a linear collider \cite{Alcaraz:2000xr}.}.

\subsection{Topcolor Confronts Precision Electroweak Observables}

In addition to the massive fermions, there is generally a heavy composite 
(predominantly of $t_L$ and $\chi_R$) scalar particle
which plays the role of the Higgs in the low energy effective theory.
For natural values of the Topcolor interaction strength
(that is, values close to the critical value required for EWSB), 
the Higgs boson mass may be computed to be roughly between 400 and 600 GeV
\cite{He:2001fz}.  One can understand this result as the value the mass
would have if the quartic coupling would become strong at the scale
$\Lambda$, about 600 GeV for $\Lambda$ of a few TeV.  

A heavy Higgs boson seems to be at variance with the measured values
of the precision electroweak observables \cite{PW}. 
However, it is also important to consider the combined radiative 
effects of the Higgs {\em and} the see-saw on the precision data, 
which can be neatly summarized in terms of the oblique correction 
parameters, $S$, $T$ (or equivalently, $\Delta \rho$), and $U$ 
\cite{Peskin:1992sw}.
The left-handed mixing between $t^\prime$
and $\chi^\prime$ leads to a modified top-quark
contribution to the $T$ parameter. The new contribution is
approximately given by replacing $m_t^2$ in the expression for $T$
with the effective value, viz.

\begin{eqnarray}
\left(m_t^{eff}\right)^2 & = & s_L^4 M_{\chi}^2 + c_L^4 m_t^2 + 
2 s_L^2 c_L^2 m_t^2 \ln\left(\frac{M_{\chi}^2}{m_t^2}\right) 
\nonumber\\
&\simeq & m_t^2 + s_L^4 M_{\chi}^2 + 2 s_L^2 m_t^2 \left[
\ln\left(\frac{M_{\chi}^2}{m_t^2}\right) - 1 \right]
\end{eqnarray}
where $M_{\chi} \simeq m_{\chi \chi}$ is the value of the 
heavy quark mass and the mixing angle $\theta_L$ is given by

\begin{equation}
s_L \equiv \sin \theta_L \simeq \frac{\mu_{t \chi}}{m_{\chi \chi}} .
\end{equation}

Thus, the model leads to a new, positive, fermionic contribution to the
$T$ parameter, which can overcome the negative
contribution induced by the presence of the heavy Higgs boson.
In contrast, the fermionic contribution to the $S$ parameter is very small, 
while the heavy Higgs boson 
(in comparison with the light Higgs seemingly favored by the
data in the absence of further ingredients) produces a
positive contribution.  The net effect is that the fermions produce
a slightly greater shift in $T$ than the shift associated with the
heavy Higgs\footnote{A similar mechanism renders the 
Topflavor Top Seesaw model \cite{He:1999vp} consistent with precision data
by mixing top and bottom with heavy vector-like weak doublets,
and predicts a Higgs mass above 200 GeV.}.
Together with the requirements that both EWSB and the correct top-quark 
mass are obtained, consistency with precision observables
further constrains the mass of $\chi$ to lie between roughly 3.8 and
7 TeV, for natural Topcolor models \cite{He:2001fz}.
Such large masses, even for strongly interacting particles, are
difficult to detect at either the LHC and a TeV linear collider.

\begin{figure}[t]
\vspace*{-1.cm}
\centerline{
\epsfxsize=10.0cm\epsfysize=10.0cm
                     \epsfbox{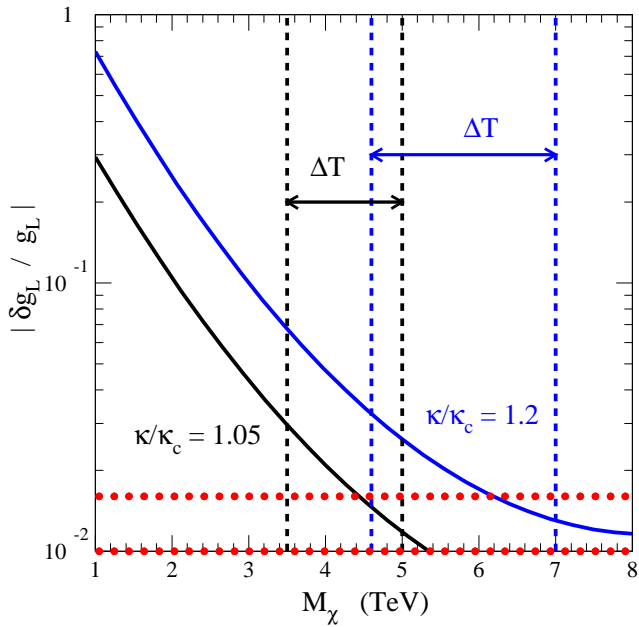}
}
\caption{\em Predictions of the Top Seesaw model with Topcolor interaction
$\kappa /\kappa_c = 1.05$ (left curves) and $\kappa /\kappa_c = 1.2$  
(right curves) for the relative shift in the $Z$-$t$-$\bar t$
interaction strength, as a function of the heavy $\chi$ mass.
Also indicated is the region of $M_\chi$ required for Top Seesaw
to produce a $\Delta T$ consistent with the precision electroweak data
(dashed curves)	and the region probed by measuring the interaction
to either 1.5\% or 1\% (dotted curves).}
\label{fig:LCtopcolor}
\end{figure}

The top coupling with the $Z$ boson, relative to its SM prediction is,
\begin{equation}
\frac{\delta g_L}{g_L} = \frac{s_L^2} 
{ 1 -  4 \sin^2\theta_W / 3} .
\end{equation}
For the viable range of parameters discussed above, the 
resulting shift in the coupling ranges between about 5\% and 1\%.
A Linear Collider will provide an accurate determination of the
top couplings to the weak vector bosons, by measuring the top
quark production at threshold. Recently, there has been great theoretical
progress in understanding this process, and it is safe to say that
the typical deviations expected in the simplest Topcolor models, as
described above, are far larger than the present theoretical
uncertainties within the SM, with the latter being presently
estimated to be less than about 1.5 percent~\cite{Hoang:2001mm}.

The values given above should be therefore be compared with the 
experimental precision on these couplings expected 
at a $\sqrt{s} = 500$ GeV
TESLA linear collider~\cite{Heuer:2001rg}, 
which are approximately 1.8 and 1.4 percent for the axial and
vector couplings, respectively. Fig.~\ref{fig:LCtopcolor} shows the 
predictions for the variation of the left handed coupling of
the top quark to the $Z$ boson as a function of the
heavy quark mass. These values are compared with the values of
the precision measurement parameter $\Delta T$ necessary to 
reproduce the precision electroweak data within the minimal
Top Seesaw model. In the region of parameters consistent with
experiment, $\delta g_L/g_L$ varies from somewhat more than
one up to about five percent. As the figure shows, such variations
are testable at a 500 GeV collider for most of the allowed
parameter space.

A TeV linear collider
will also be able to discover the heavy Higgs boson. The
determination of the Higgs boson mass together with the precise
measurements of the top quark couplings to the weak gauge
bosons would serve to identify the two key ingredients of the
Top Seesaw mechanism of EWSB: namely, a heavy Higgs boson,
and extended interactions of top.  
The knowledge of the Higgs mass and top
coupling deviations would provide a prediction for a range
of $\chi$ masses compatible with the Top Seesaw model.
Therefore, a TeV Linear Collider will become a Top Seesaw analyzer.

\section{Two Higgs Doublet Model}
\label{sec:2hdm}

A second example of a model in which the precision electroweak 
observables may be satisfied, even in the presence of a heavy
SM-like Higgs boson, is a two Higgs doublet model. 
As we explained in the introduction, in this model, the contribution to the
parameter $\Delta \rho$ accruing from the second Higgs doublet
may be compensated by the one from the heavy SM Higgs, bringing
the predictions for the precision electroweak observables back
within the experimental $1 \sigma$ contour. 

A generic two Higgs doublet model, of course, is characterized by quite a few 
parameters. Of particular interest to us are $\tan \beta$ (the ratio 
of the two vacuum expectation values), the Higgs mixing angle $\alpha$, 
and the four masses: $m_A$ (the pseudoscalar), $m_{H^\pm}$ (the charged
Higgs) and those for the light and heavy CP-even neutral states namely
$m_h$ and $m_H$. Along with the common mass parameter 
($\mu_{12}^2 \phi_1^\dagger \phi_2$), these determine all the 
properties of the Higgs including their self interactions. 

The relevant question for such models relate to the observability 
of at least one of the non-standard Higgs bosons (along with the 
SM-like Higgs) at a TeV linear collider. 
If the mass of the SM-like Higgs boson is close to the present
experimental bound on this quantity, $m_{H^{SM}} \geq$ 114 GeV,
then the non-standard Higgs bosons can obviously be arbitrarily heavy,
provided they are sufficiently degenerate.
On the other hand, if the mass of the SM-like Higgs boson is
large, 
\begin{equation}
M \equiv
m_h^2 \; \sin^2(\alpha - \beta) + m_H^2 \; \cos^2(\alpha - \beta) 
\geq m_{\rm bound}^2 ,
\label{eq:SMHm}
\end{equation}
with $m_{\rm bound}$ larger than about 400 GeV, then the extra contributions
to the precision electroweak observables demand either light
non-standard Higgs bosons, observable at a TeV Linear
Collider, or very large mass splittings between the second-Higgs doublet
components.  This can be understood by noting that, for mass differences 
$\Delta m^2$ smaller
than the average mass $m^2$, the contributions to the parameter
$\Delta \rho$ are proportional to
\begin{equation}
\Delta \rho \simeq \frac{ \left(\Delta m^2 \right)^2 }{ m^2 }
\end{equation}
and therefore, the larger $m^2$ is, the larger $\Delta m^2$ should be
in order to produce the same effect. 

The precise one-loop contributions to the parameters $S$ and $T$ are
given by~\cite{Inami:1992rb,Chankowski:2000an,He:2001tp}

\begin{eqnarray}
\Delta \rho & = & \frac{\alpha}{16 \pi \sin^2\theta_W m_W^2}
\left\{
\cos^2(\beta - \alpha) \left[ f(m_A,m_{H^{\pm}}) + f(m_{H^{\pm}},m_h)
- f(m_A,m_h) \right] \right.
\nonumber\\
& + & \left. 
\sin^2(\beta - \alpha) \left[ f(m_A,m_{H^{\pm}}) + f(m_{H^{\pm}},m_H)
- f(m_A,m_H) \right] \right\}
\nonumber\\
& + & \cos^2(\alpha - \beta) \Delta \rho_{SM}(m_{H}) +
\sin^2(\alpha - \beta) \Delta \rho_{SM}(m_h) - \Delta \rho(m_{H^{SM}})
\end{eqnarray}
with
\begin{equation}
f(x,y) = \frac{x^2 + y^2}{2} - \frac{x^2 y^2}{x^2 - y^2} 
\log \frac{x^2}{y^2}
\end{equation}
and 
\begin{equation}
\Delta \rho_{SM} = \frac{ 3 \alpha}{16 \pi \sin^2\theta_W m_W^2}
\left[ f(m,m_Z) - f(m,m_W) \right] - \frac{\alpha}{8 \pi \cos^2 \theta_{W}}
\end{equation}
and where we have subtracted the SM contribution. Similarly,
\be
\barr{rcl}
S & = &   \dis \frac{1}{12 \pi}
\Bigg\{ \cos^2(\beta - \alpha ) \left[ 
\log \frac{m_H^2}{m_{H^{SM}}^2} + \log \frac{m_h m_A}{m_{H^{\pm}}^2}
\right.  \\[2ex]
& & \dis \hspace*{3em} + \left. \left .
2 \frac{m_h^2 m_A^2}{\left(m_h^2 - m_A^2\right)^2} +
\frac{\left(m_h^2 + m_A^2\right)\left(m_h^4 + m_A^4 - 4 m_h^2 m_A^2\right)}
{\left(m_h^2 - m_A^2 \right)^3 }\log\frac{m_h}{m_A} \right]
\right.
\\[2ex]
& & \dis \hspace*{3em} + 
\sin^2(\beta -\alpha) \left[ (m_h \leftrightarrow 
m_H \right)] - \frac{5}{6} \Bigg\}
\earr
\ee

Now, large 
splittings between the masses of the 
different non-standard Higgs bosons can only
be obtained via large values of some of the quartic couplings. Similarly, 
a large value for the effective SM-like Higgs mass (eq. \ref{eq:SMHm})
is again obtained only through large values of the quartic couplings. 
Thus, the requirement
of perturbative consistency of the model up to some high energy scale
$\Lambda_{\rm new}$ serves to put constraints on the Higgs 
masses\footnote{A similar statement also holds for the one doublet 
	case (SM).}.
Since we are interested in the case of a heavy Higgs
boson, with mass ${\cal{O}}$(500 GeV), we should demand 
that the theory be perturbative up to at least a few TeVs. 

The requirement of perturbative consistency of the theory has two
important effects. On the one hand, it sets a limit on the SM-like Higgs
boson mass. On the other, it sets a limit on the possible 
splitting of masses of the non-standard Higgs masses and therefore
on the average mass of the non-standard Higgs bosons for a given value
of their contribution to the precision electroweak observables. 

The bound on the SM-like Higgs boson is enhanced for smaller values 
of the cutoff scale. The bound on the non-standard Higgs boson masses
is correlated with the exact value of the Higgs boson mass within
the allowed range: the larger the SM-like Higgs boson mass, the larger the
necessary new contributions to the precision electroweak observables.
For a given value of the average mass of the non-SM Higgs boson,  these
large contributions can only be obtained via larger quartic couplings
and, since these are restricted by the perturbative bound,
a stricter upper bound on the non-standard Higgs boson masses is obtained.

\begin{figure}[htb]
\vspace*{-4.cm}
\centerline{
\epsfxsize=9.cm\epsfysize=12.0cm
                     \epsfbox{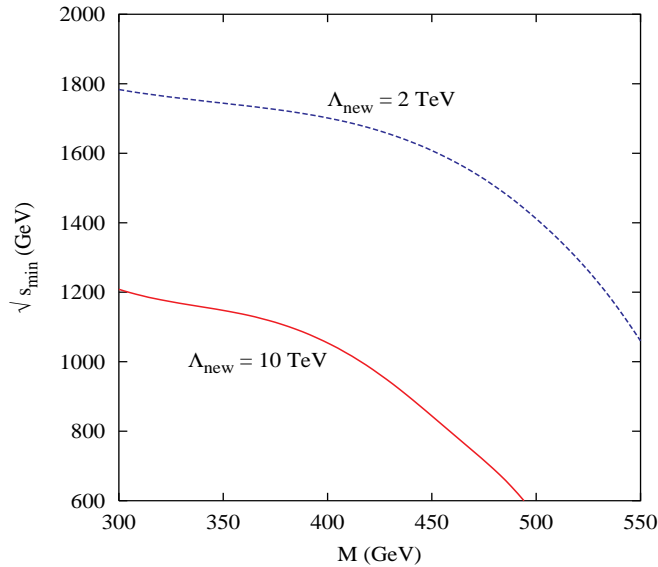}
}
\caption{\em The minimum center-of-mass energy an $e^+ e^-$ linear collider
would need in order to observe at least one non-standard Higgs through
production of $H^+ H^-$, $H A$, $h A$, $h Z$, or $H Z$, as a function of
the effective SM-like Higgs mass and demanding perturbativity up to 
2~TeV (upper curve) or 10~TeV (lower curve).
	}
\label{fig:LC2HDM}
\end{figure}

In Fig.~\ref{fig:LC2HDM}
we show, for $\Lambda_{\rm new} = 10$ TeV  and $\Lambda_{\rm new} = 2$ TeV,
the necessary center of mass energy of a linear collider to ensure
the observability of at least one non-standard Higgs boson, as 
a function of the SM-like Higgs boson mass. 
To obtain this figure we demand that
 \begin{itemize}
\item The values of the predicted $S$ and $T$
	parameters lie within the 90\% C.L. ellipse 
	about the experimentally measured values;
\item None of the five quartic couplings should become nonperturbative 
      (under one-loop renormalization group evolution) 
      at any scale below $\Lambda_{\rm new}$ for $\alpha_S(M_Z) = 0.118$
      and $m_{top} = 175$ GeV;
\item Since the theory is only an effective one 
      below $\Lambda_{\rm new}$, the masses should all be much lighter  
      than $\Lambda_{\rm new}$; 
      we use the rather loose criterion suggested by 
      Hasenftatz {\em et al.}~\cite{Hasenfratz:1987tk}, namely
      $M^2 < \Lambda^2_{\rm new} / 2$ for all particles
      in the spectrum;
\item To be considered observable, the 
      Higgs bosons must be produced in pairs ($H^+ H^-$, $H A$, $h A$) 
      or in conjunction with a gauge boson ($h Z$, $H Z$) with a cross
      section larger than 0.1 fb (corresponding to 50 events with
      500 ${\rm fb}^{-1}$).
\end{itemize}
The last condition is somewhat heuristic, and would be improved by detailed
simulation of each production mode, including Higgs decays and backgrounds.  
Such details for the exotic Higgses are highly model-dependent, and 
beyond the scope of this work.

For values of the Higgs boson mass somewhat larger than 200 GeV,
a non-trivial bound on the non-standard Higgs bosons develops,
provided the only source of new physics with a relevant impact on
precision electroweak observables is an additional Higgs doublet. 
This bound is quite weak for small values of the SM-like Higgs boson mass, 
but becomes stronger as this mass increases. For the values
of the cutoff scale considered, $\Lambda_{\rm new} \geq 2$ and 10 TeV, 
the effective Higgs boson mass $M$ cannot be larger than about 
550 and 500 GeV, respectively. Therefore,
the first conclusion obtained from such considerations is
that at least one of the Higgs bosons will
be observable at a TeV Linear Collider. The second conclusion,
which can be extracted from the results displayed in Fig.~\ref{fig:LC2HDM}
is that for $\Lambda_{\rm new} \simgt 10$ TeV and SM Higgs boson masses 
larger than 300 GeV, a linear
collider with center of mass energy of about 1.2 TeV will be 
enough to ensure the observability of at least one non-standard
Higgs boson.  Observe that for $M \simgt$ 450 GeV, both CP even
Higgses may be produced in association with a $Z$ boson, implying
values of $\sin^2(\alpha - \beta)$ significantly different from one.

It might be be argued that the fourth requirement imposed in obtaining
Fig.~\ref{fig:LC2HDM} is unnecessarily restrictive. The 
Higgs particles could as well be 
produced in association with fermions, particularly the bottom 
and the top. As is well known, such couplings may be significantly 
enhanced over their SM counterparts. However, we have desisted from 
considering these processes, simple as they are, in view of the additional
model dependence. The results of Fig.~\ref{fig:LC2HDM} are thus 
{\em conservative} and the actual value of $\sqrt{s}_{\rm min}$ may be 
significantly smaller within a given scenario of Higgs-fermion couplings.

\section{Beautiful Mirrors}
\label{sec:bmirrors}

Until now, we have discussed the constraints coming from the
precision electroweak data in terms of the oblique parameters
S, T and U. The LEP and SLD precision electroweak measurements
are also sensitive to non-oblique corrections. The most important
one of these is related to the couplings of the left-
and right-handed bottom quark
to the neutral gauge boson. In particular, measurements of 
the partial hadronic width of the Z boson decays into bottom
quarks and of the forward-backward and 
left-right asymmetries of bottom quarks at
the Z peak have been obtained, which have led to a precise
measurement of the bottom quark couplings in terms of the
electroweak mixing angle. 

The forward-backward bottom quark asymmetry measured at LEP ($A^b_{FB}$)
presents a 2.9 $\sigma$ deviation from the best fit value
within the SM. Such a large deviation becomes
particularly significant, since it pushes the
Higgs boson mass toward values consistent with the constraints
coming from direct searches. Indeed, if the hadronic asymmetries
were ignored, the best fit to the Higgs boson mass would
lead to a value which is well below the present experimental
bound. Probably equally significant is the fact that if
this deviation is to be explained by a modification of the
effective couplings of the bottom quarks to the $Z$
boson, the absolute value of the coupling of the right handed 
bottom quark should be 26 percent larger than the value predicted
in the SM. Such a large deviation can be 
obtained in a natural way via tree-level mixing of the bottom quark 
with new, heavier quarks, with the same charge and color quantum
numbers as the bottom quark.

In Ref.~\cite{Choudhury:2001hs}, we performed an analysis of the possible
impact on the fit to the electroweak precision data proceeding
from the presence of additional vector-like quarks in the
spectrum. We showed that a significant improvement in the fit
to the precision electroweak observables could be obtained
by demanding a correlation between the Higgs boson and the
new quark masses. Moreover, we showed that the introduction of
the new degrees of freedom may lead to an improvement of the
condition of unification of couplings, at a sufficiently large
scale in order to avoid a conflict  with proton decay constraints.

The details of these models are in Ref.~\cite{Choudhury:2001hs}.
Here we concentrate on one of the two possibilities 
studied in \cite{Choudhury:2001hs}, which leads to the presence of a heavy 
Higgs boson in the spectrum. In this case, apart from the heavy Higgs
boson, with masses in the range 250 to 500 GeV, one encounters
a new doublet of quarks and its mirror partner, with masses
lower than 300 GeV.  The doublet contains a new top-like quark ($\chi$)
and a bottom-like quark ($\omega$), with vector-like weak interactions.
The right-handed $\omega$ mixes with
the right-handed bottom quark, with a large mixing angle
\begin{equation}
\sin^2\theta_R \simeq 0.3 ,
\end{equation}
resulting in predictions for $A^b_{FB}$ consistent with measurements.

The new bottom and top quarks will be produced and observed
at the Tevatron collider. However, a hadron collider will
have difficulty measuring the bottom mixing angle, which is only relevant
for the weak interactions, and is thus likely to be swamped by large
hadronic backgrounds.  A linear collider
can directly measure the mixing by observing the flavor violating process:
\begin{equation}
e^+e^- \to Z^* \to \overline{b} \omega
\end{equation}
which is directly proportional to the mixing parameters 
$\sin^2 \theta_R \cos^2 \theta_R$.  The dominant decay mode of the
$\omega$ is also through this mixing, $\omega \ra Z b$ with a branching
ratio of nearly one.  For a sufficiently heavy Higgs boson, decay modes
involving the beautiful mirrors such as 
$H \ra \overline{\omega} b \; (\overline{b} \omega)$ may be allowed.  For the
relevant model parameters consistent with precision electroweak data, the
partial widths into these modes are less than the width into top quarks.
Therefore, the branching ratio of Higgs into gauge bosons and top quarks will
generally differ from the SM predictions by about $10\%$.

\begin{figure}[htb]
\vspace*{-1cm}
\centerline{
\epsfxsize=10.0cm \epsfysize=10.0cm
                     \epsfbox{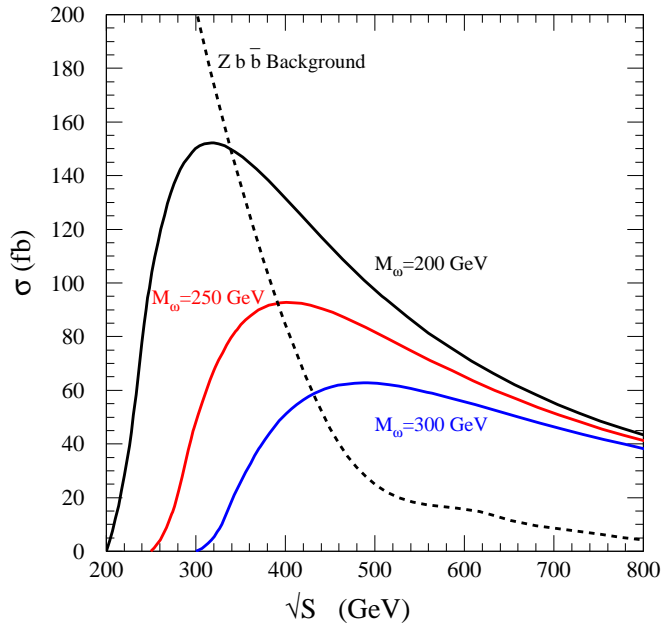}
}
\caption{\em Cross sections for $e^+ e^- \ra \omega \overline{b}$ (
$\overline{\omega} b$) production as a function of collider energy (solid
curves), for three $\omega$ masses consistent with the beautiful mirrors
explanation of the LEP data.  The dashed line indicates the background
from $e^+ e^- \ra Z b \overline{b}$ (including a Higgs of mass 300 GeV).
All cross sections are for unpolarized beams and the cuts described in the
text.}
\label{fig:bomega}
\end{figure}

In order to estimate
how well a linear collider could measure $\sin\theta_R$ through this
process, we simulate both signal and background (including a
300 GeV Higgs boson appropriate for this 
model\footnote{While we choose to display 
	our results for this particular value of the Higgs mass, 
	the conclusions are quite robust and do not change 
	qualitatively for $250 \: {\rm GeV} \lsim m_H \lsim 600 \: {\rm GeV}$.
	}) and apply simple 
kinematic cuts to simulate the detection efficiency of the bottom quarks, 
$p_T^b \geq 15$ GeV and $|y^b| \leq 3$.  
One can further reduce the background by applying a very loose cut on the 
$\omega$ mass, requiring that the invariant mass of the $Z$ with one or the
other of the bottoms is within 50 GeV of $m_\omega$.  
We do not assume a specific decay mode of the $Z$ boson.
In Fig.~\ref{fig:bomega}, we present the resulting signal cross sections
for three values of $m_\omega$ and the background after applying the
cuts described above.  Comparison of the signal rates with the background
indicates that signal rates equal to or greater than the background
are obtained for collider energies greater than roughly 
$m_\omega + 100$ GeV, with cross sections greater than 50 fb .  
Thus the entire parameter space favored by precision data can be studied
at a 500 GeV linear collider.  

We have chosen a large invariant mass window of $m_\omega \pm 50$ GeV,
which Fig.~\ref{fig:bomega} indicates is sufficient to discover 
the flavor-violating signal.  The physical width of the $\omega$ 
is of the order of a few GeV, and thus once the signal is found, the
cut on the invariant mass may be tightened.  Once it is less than 20 GeV,
still much larger than the expected jet resolution,
the background becomes essentially zero at all energies, and the signal
remains unchanged.  We estimate the precision of the measurement of the 
cross section from its statistical uncertainty and find that 
with 100 ${\rm fb}^{-1}$ we can measure 
$\sin^2 \theta_R$  at an accuracy of $2.3\%$ when $\sqrt{s}=500$ GeV
and to $2.6\%$ when $\sqrt{s}=800$ GeV for all $\omega$ masses below 300 GeV.
Together with the observation of the heavy Higgs,
this would establish quite precisely the beautiful mirror interpretation of
present-day electroweak data.

\section{Conclusions}
\label{sec:conclusions}

Precision electroweak observables suggest
the presence of a light SM-like Higgs boson, with mass
close to the present experimental limits.
The study of such a light Higgs is one of the primary motivations for a 
TeV scale linear collider.
However, the nature of indirect constraints is such that 
the simplest explanation need not be the correct one, and thus it
behooves us to consider the alternative that the Higgs is heavy.
Models with heavy Higgs bosons are able to obtain consistency with
the data by introducing new contributions to the
precision electroweak observables, which compensate
the additional contribution induced by the heavy 
SM-like Higgs, restoring the agreement between the
theoretical predictions and the experimental results.

From the phenomenological point of view, the analysis
of the signatures associated with a heavy Higgs presents
interesting challenges, since it will be associated with
a broad resonance that cannot be described by the simple
production and decay signatures usually used in the narrow
width approximation. We have presented possible ways of
disentangling these signatures at a linear collider and
provided an estimate of the possible reach of a 800 GeV
linear collider in the most important channels.

In this article, we have also studied some the most natural
models leading to the presence of a heavy SM-like
Higgs boson in the spectrum: the simplest Topcolor model, 
a two Higgs doublet extension of the SM
and an extension with vector-like quarks
which lead to an improvement in the fit of the 
measured couplings of the bottom quark to the $Z$
gauge boson. In the first case,
we show that, although the new physical degrees of freedom
may remain beyond the reach of even the LHC, the linear collider
has the potential of testing these models via the measurement
of the Higgs mass and the coupling of the top quark to the
massive weak gauge bosons. In the second case, we have
shown that, if we demand perturbative consistency
of the theory up to a few TeV scale, at least one Higgs boson
should appear in the spectrum and, if the SM-like 
Higgs boson is heavy, with mass larger than 500 GeV, at least
one non-standard Higgs boson should appear at the reach of 
a TeV  linear collider. Finally, in the model with extra
light vector-like quarks, a precise measurement of the associated
flavor violating processes will lead to an accurate measurement
of the non-trivial mixing of these quarks with the third generation
quarks and therefore provide a crucial test of this kind of model.

\newpage
~\\
{\Large \bf Acknowledgements}\\
~\\
The authors are grateful for conversations with S. Mrenna and C.--P. Yuan.
Work at ANL is supported in part by the US DOE, Div.\ of HEP,
Contract W-31-109-ENG-38.
DC thanks the the HEP Division of the Argonne National Lab. 
and the Theory Division of Fermilab for hospitality while 
part of the project was being carried out and 
the Deptt. of Science and 
Technology, India for financial assistance under the 
Swarnajayanti Fellowship grant.

\end{document}